\documentclass[a4paper,usenatbib]{mnras}

\usepackage{graphicx,natbib,hyperref,amsmath} % Required for inserting images
\usepackage{newtxtext,newtxmath}
\usepackage{xcolor}
\usepackage{anyfontsize}

\newcommand{\diffhydro}{\texttt{diffHydro}}

\title[Differentiable Cosmological Hydrodynamics]{Differentiable Cosmological Hydrodynamics for Field-Level Inference and High Dimensional Parameter Constraints}
\author[B. Horowitz and Z. Luki\'c]{
Benjamin Horowitz$^{1,2,3}$\thanks{E-mail: ben.horowitz@ipmu.jp} and
Zarija Luki\'c$^{3}$
\\
$^{1}$Kavli IPMU (WPI), UTIAS, The University of Tokyo, Kashiwa, Chiba 277-8583, Japan \\
$^{2}$Center for Data-Driven Discovery, Kavli IPMU (WPI), UTIAS, The University of Tokyo, Kashiwa, Chiba 277-8583, Japan\\
$^{3}$Lawrence Berkeley National Lab, 1 Cyclotron Road, Berkeley, CA 94720, USA
}
\date{January 2025}

\begin{document}
\maketitle

\begin{abstract}
Hydrodynamical simulations are the most accurate way to model structure formation in the universe, but they often involve a large number of astrophysical parameters modeling subgrid physics, in addition to cosmological parameters. This results in a high-dimensional space that is difficult to jointly constrain using traditional statistical methods due to prohibitive computational costs. To address this, we present a fully differentiable approach for cosmological hydrodynamical simulations and a proof-of-concept implementation, \diffhydro. By back-propagating through an upwind finite volume scheme for solving the Euler Equations jointly with a dark matter particle-mesh method for Poisson equation, we are able to efficiently evaluate derivatives of the output baryonic fields with respect to input density and model parameters. Importantly, we demonstrate how to differentiate through stochastically sampled discrete random variables, which frequently appear in subgrid models. We use this framework to rapidly sample sub-grid physics and cosmological parameters as well as perform field level inference of initial conditions using high dimensional optimization techniques. Our code is implemented in JAX (python), allowing easy code development and GPU acceleration. 

\end{abstract}

\section{Introduction}

A core goal of modern cosmology is to understand the formation of the large scale structure of the universe and trace its evolution over cosmic time. However, most of the mass-energy density of the universe is found in the dark sector (i.e. dark matter and energy) and we are left interpreting is impact on baryons as biased tracers of this underlying distribution. This presents major theoretical and practical challenges to calculate and propagate uncertainties to constrain cosmological models.
%\kg{Conversely, the most visible manifestation of the cosmic baryons are in stars and galaxies, the growth and evolution of which are one of the largest questions in contemporary astrophysics.}

A key tool for modeling the baryons together with the dark sector is cosmological hydrodynamical simulations \citep{1992ApJS...78..341C}. While the dark matter can be modeled as a collisionless fluid via pure gravitational dynamics, baryonic gas is collisional and therefore requires solving the compressible Euler fluid equations. 
In addition, small-scale astrophysical phenomena, such as star formation and AGN activity \citep{2006PhFl...18h5101D,2008ApJ...680.1083R,2012MNRAS.422.2609R,2013MNRAS.429.3068T,2022ApJS..262....9O,2024ApJ...975..183O} operate on scales not directly resolvable in large-scale cosmological simulations, ``subgrid'' models approximate their influence on the larger, resolved structures. Jointly solving this evolution is computationally demanding and requires the introduction of additional modeling parameters beyond the cosmological model itself. Relevant summary statistics are then calculated from outputs of these hydrodynamical simulations and compared to the observational data from sky surveys \citep{2023MNRAS.526.6103K}.

However, a full likelihood analysis requires exploration of the entire parameter space of cosmological models. This usually entails running many hydrodynamical simulations sampling a model parameter space, with either summary statistics interpolated in this space \citep[e.g.~power spectra emulators][]{Heitmann2010,2019ApJ...872...13W,2021JCAP...12..046G} or a machine learning-based augmentation \citep{2019Horowitz,2022ApJ...929..160H,2024arXiv240115891S}. Depending on the dimensionality of the parameter space and the cosmic volumes under consideration, this can be extremely expensive computationally. To simplify this, many astrophysical parameters are fixed via manual calibration via various astrophysical observables (i.e. galaxy luminosity functions, M-$M_{*}$ relations, etc.), many of which also have some cosmological dependencies \citep{2017MNRAS.465.3291W}.

In recent years, there has also been an increased interest in moving beyond summary statistics and to use field level inference for cosmological constraints and cosmic structure inference \citep{2013MNRAS.432..894J,2017JCAP...12..009S,2022NatAs...6..857A,2024arXiv240303220N}. These methods use full forward modeling from the initial conditions of the universe to the late time observable field to then compare with data. This requires performing a high dimensional optimization and/or sampling algorithm, relying on derivative back-propagation through the dynamical model \citep{2017JCAP...12..009S,2022arXiv221109958L}. Perturbation and particle mesh methods for gravity evolution can be differentiated straightforwardly, and then mapped to the the observable signal via bias models \citep[e.g.][]{2019TARDISI,2021TARDISII}. 

Uncertainties in this mapping, however, significantly limit the constraining power. For example, there are dense proto-clusters at $z\sim2.3$  with similar galaxy number densities but vastly different Lyman-$\alpha$ forest statistics \citep{2023ApJ...945L..28D,2024MNRAS.532.4876D} in ways that are difficult for analytical prescriptions to capture \citep{2022constrainfgpa}. One approach to these modeling issues is by emulation of the baryonic fields via machine learning algorithms \citep{2018JCAP...10..028M,2022ApJ...929..160H,2022hyphy}. In this way, a model is trained based on simulations to map from a dark matter field to the baryon fields of interest. However, this relies on calibration against a large set of simulations spanning both cosmological and astrophysical parameters. A full analysis may involve tens if not hundreds of parameters, requiring a prohibitively large number of simulations to span the plausible parameter space and train an emulator.

In this work, we present an alternative approach; we make the hydrodynamical simulation itself differentiable with respect to the initial conditions and parameters, both cosmological and stochastic astrophysical. This enables a wide range of analysis that are otherwise impossible. This includes; rapid tuning of sub-grid physics parameters, joint calibration of astrophysical and cosmological parameter from observed summary statistics, and field level inference of hydrodynamical fields/related observables.

By leveraging existing automatic differentiation machinery developed for deep neural networks \citep{2016arXiv161100712M,2021ascl.soft11002B}, our implementation has comparable computational cost and memory usage as a standard GPU-accelerated hydrodynamical simulation or large machine learning model. We emphasize, however, that there is no emulation or calibration within this method and we directly solve for all physics as described below. In addition, our implementation is based on the \texttt{JAX} framework, which utilizes Just-In-Time compilation, enabling extremely fast function evaluations on a GPU. \texttt{JAX} has been used for differentiable simulations in a wide range of physical simulations outside astrophysics, including Boltzmann lattice models for fluids \citep{ataei2024xlb} and plasma physics in tokamak reactors \citep{torax2024arxiv}. Our software, \diffhydro\, retains high user-readability, allowing rapid development and a useful test-bed for new numerical routines, subgrid prescriptions, and observed fields.

For forward modeling gas dynamics, we implement a total variation diminishing (TVD) method for solving Euler equations. TVD methods are widely used for solving the Euler equations as they effectively capture shock waves and discontinuities while minimizing unphysical oscillations. These methods offer high-resolution solutions in smooth regions and maintain numerical stability due to their ability to limit total variation \citep{Shu1988, Harten1997}. They also conserve mass, momentum, and energy, which is essential for accurate fluid flow simulations \citep{LeVeque2002}.  While they perform well with shocks, they are less effective at handling contact discontinuities and may introduce over-dissipation, reducing the sharpness of shock waves \citep{Laney1998}.

We want to emphasize that modern cosmological hydrodynamical simulations \citep[e.g.][]
{2014MNRAS.445..175G,2015ApJS..217...24S} employ complex numerical methods and our implementation is a proof of principle. While we demonstrate the key components of differentiable hydrodynamical simulations and how they can be used for field-level inference or to constrain parameters of interest, more development is needed to apply these methods at scale for ongoing and upcoming cosmological surveys.

\section{Methods}

In this section, we briefly describe the equations we use to solve the underlying gas and dark matter evolution in \diffhydro.

We are particularly focused on a formalism that is differentiable; i.e. we want to be able calculate derivatives of the resulting simulation fields, i.e. density, internal energy, velocity, etc., with respect to both the model parameters, such as cosmological parameters, as well as the initial conditions of the simulations. If this is computationally tractable, we can further calculate statistics, such as power spectra, from those fields and use the chain-rule to calculate the derivative of those statistics with respect to the astrophysical/cosmological parameters and initial conditions. 

\subsection{Initial Conditions}
\label{subsec:ics}

To set our initial conditions, we use the Einsten and Hu fitting formula for the matter transfer function \citep{1999ApJ...511....5E,1998ApJ...496..605E} as implemented in the \texttt{JAX-COSMO} package \citep{2023OJAp....6E..15C}. Since this formula is analytic, we can differentiate our model with respect to cosmological parameters and deviations from this model are not expected to be significant at scales of interest for our analysis.  We then use LPT to evolve the initial gaussian realization to $z=99$ to set our initial conditions for the full hydrodynamical simulation.

An alternative approach is to take the underlying Boltzmann codes (i.e. CLASS \citep{2011arXiv1104.2932L} or CAMB  \citep{2011ascl.soft02026L}) and make it differentiable. This could be done either by direct backpropagation through the solver \citep{2024JCAP...06..063H,2023zndo..10065126L} or via precomputed emulation \citep{2023ascl.soft07010A}. The relative importance of the full treatment depends on the end application.

\subsection{Dark Matter}
\label{subsec:dm}

Current cosmological constraints \citep{2020A&A...641A...6P} indicate that the matter content of the universe is dominated by a non-relativistic, pressure-less fluid known as cold dark matter. The evolution of the phase space distribution function, $f$, of this fluid in an expanding space-time can be described by the Vlasov equations,
\begin{equation}
\label{eq:dm_vlasov}
\frac{\partial f}{\partial t} + \frac{1}{ma^2} \mathbf{p} \cdot \nabla f 
                             - m \nabla \phi \cdot \frac{\partial f}{\partial \mathbf{p}} = 0 \, \, , 
\end{equation}
where $a$ is the cosmological scale factor, $\mathbf{p}$ is the momenta, $m$ is the particle mass, and $\phi$ is the gravitational potential. In practice solving this as a continuum equation is impractical in the cosmological context and instead the common approach is to monte carlo sample it at some initial time. We then evolve the discrete particles as an N-body system. Hamilton's equations for this system become
\begin{eqnarray}
\label{eq:dm_hamil}
    \frac{d \mathbf{x}_i}{d t}       = \frac{1}{a} \mathbf{u}_i \\
    \frac{d (a \mathbf{u}_i)}{d t} = \mathbf{g}_i
\end{eqnarray}
where $\mathbf{x}$ ($\mathbf{u}$) is the comoving location (peculiar velocity) for $i$th particle and $\mathbf{g}_i$ is the gravitational field evaluated at $\mathbf{x}_i$ at time $t$.

The dark matter evolution requires special care in order to allow easy propagation of gradient information. Standard N-body codes that rely on calculating particle-particle interactions described in Eq. \ref{eq:dm_hamil} on small scales, \citep[see, for example][]{2001gadget,2002JApA...23..185B,Habib2016} rely on constructing large tree structures \citep{1986Natur.324..446B} for calculating forces on particles in a computationally efficient way. However these structures pose significant challenges for back-propagation due to the discrete assignment into the tree structure. %\zl{This requires some clarification...} 
In particular, tree-based methods are generally piece-wise constant and therefore their derivatives are zero almost everywhere. While certain graph theory algorithms can be made differentiable \citep{10.5555/3495724.3496521}, the memory requirements for these approaches generally scale poorly with particle number. In addition, for the application of field level inference, actual reconstruction of individual particle orbits (i.e. in virialized systems) is likely both unnecessary for cosmological observables and unwanted due to the introduction of significant posterior degeneracies due to chaotic orbits.% \zl{particle trajectories are formally chaotic in halos...}

Instead we utilize a particle-mesh scheme \citep{1988csup.book.....H,1991ApJ...368L..23C}, in particular a \texttt{JAX} implementation \citep{2022mla..confE..60L} of a fast particle mesh solver \citep{2016MNRAS.463.2273F}, which utilizes a modified kick-drift-kick scheme for particle evolution \citep{1997astro.ph.10043Q}. As forces are calculated via a potential formalism, we are able to do calculations on a grid with standard linear algebra and Fourier operators. This allows straight-forward backpropagation via automatic differentiation. 

We note that using a Particle Mesh scheme without adaptive refinement fundamentally limits ability to resolve small scale dark matter structures. Depending on the physics of interest, this may pose a fundamental limitation in the ability of this particular approach. Certain techniques, including Neural-ODE solvers for integration \citep{2022mla..confE..60L,2023A&A...679A..61L,2023arXiv231118017P} or potential gradient descent methods \citep{2018JCAP...11..009D}, have been demonstrated to improve results and significantly reduce differences between PM and full N-body algorithms.

\subsection{Gas Physics}
\label{subsec:gas}

We solve Euler equations of gasdynamics in a coordinate systems comoving
with the cosmological expansion. Comoving density and pressure of the gas are related to the proper density via the cosmological scale factor, $a$, as $\rho_b = a^3 \rho_{\rm prop}$ and $p_b = a^3 p_{\rm prop}$. We can then write the continuity and momentum equation for the comoving density/pressure in terms of peculiar baryon velocity, $U$, as 
\begin{equation}
    \label{eq:cont}
    \frac{\partial \rho_b}{\partial t} = - \frac{1}{a}\nabla \cdot (\rho_b U),
\end{equation}
\begin{equation}
    \label{eq:momentum}
    \frac{\partial (a \rho_b U)}{\partial t} = - \nabla \cdot (\rho_b UU) - \nabla p + \rho_b \textbf{g},
\end{equation}
where the gravitational acceleration is defined as $\textbf{g} = - \nabla \phi$, where $\phi$ is the gravitational potential.

For now, we use a single energy formalism where we keep track of the total energy, $E$, in each cell, which can be related to the internal energy, $e$, as $E = e + \frac{1}{2}U^2$. This can straightforwardly be expanded to a dual energy formalism, bringing more accurate solutions in low-density, high-velocity regions of the flow. The evolution of the total energy is given as;
\begin{equation}
    \frac{\partial (a^2 \rho_b E)}{\partial t} = -a \nabla \cdot (\rho_b U E + p U) + a\rho_b U \cdot \textbf{g} + a \dot{a} ((2-3(\gamma -2))\rho_b e).% + a \Lambda_{HC},
\end{equation}
Baryonic matter is generally assumed to follow a power-law equation of state of the form $p_b = (\gamma - 1) \rho e$, with $\gamma = 5/3$. This allows us to simplify the total evolution evolution as
\begin{equation}
    \frac{\partial (a^2 \rho_b E)}{\partial t} = -a \nabla \cdot (\rho_b U E + p U) + a(\rho_b U \cdot \textbf{g}),% + \Lambda_{HC}).
    \label{eq:ev}
\end{equation}

We note that there is often a heating/cooling term as well, $\Lambda_{HC}$, which we are not including in this work in order to better elucidate our simulations performance without confounding factors. This term is straightforward to include in our framework via a pre-computed lookup table.

More generally, we can re-express Equations \ref{eq:cont}, \ref{eq:momentum}, and \ref{eq:ev} in vector form in terms of $\textbf{U}=(\rho_b,a \rho_b U_x,a \rho_b U_y,a \rho_b U_z, a^2 \rho_b E)=(\rho_b,a \rho_b U, a^2 \rho_b E)$, via the following combined conservation equation,
\begin{equation}
    \frac{\partial \textbf{U}}{\partial t} = - \nabla \cdot \textbf{F} + S_e + S_g, %+ S_{HC}
    \label{eq:conservation}
\end{equation}
where our flux vector is  $\textbf{F} = ((1/a) \rho_b U, \rho_b UU, a(\rho_b U E + p U))$, the internal energy evolution source is $S_e = (0,0,-ap\nabla \cdot U)$, and the gravitational source term is $S_g = (0,\rho_b \textbf{g},a\rho_b U\cdot \textbf{g})$.  %double check source terms? Not certain about S_e

\subsubsection{Relaxed Total Variation Diminishing Method}

%We adopt a monotone upwind schemes for conservation laws, specifically a half time-step up-winding followed by a full step second order total variation diminishing (TVD) scheme \citep{1983JCoPh..49..357H}. 

We solve these equations on a fixed grid, with our state vector $\textbf{U}$ defined in the center of each grid cell and our fluxes defined on the boundaries between cells. We use a relaxation method \citep{jin1995relaxation}, which has been shown to be effective at resolving astrophysical shocks \citep{1998ApJS..115...19P}. This approach follows the procedure described in \citet{2003PASP..115..303T}.

In one dimension, we can adopt a symmetric approach to general advection by decomposing our state vector into a left moving and right moving flow, i.e.
\begin{equation}
    \textbf{U}^R =  \left( \frac{1+v/|v|}{2}\right) \textbf{U},
\end{equation}
\begin{equation}
    \textbf{U}^L =  \left( \frac{1-v/|v|}{2}\right) \textbf{U}.
\end{equation}
We can rewrite Eq. \ref{eq:conservation} in terms of the left and right moving fluxes as
\begin{equation}
    \frac{\partial \textbf{U}}{\partial t} + \frac{\partial \textbf{F}^R}{\partial t} -  \frac{\partial \textbf{F}^L}{\partial t} = 0,
\end{equation}
where the fluxes are defined as $\textbf{F}^i = c\textbf{U}^i$ where $c = |v| + c_s$ is the ``freezing speed" with $c_s$ the speed of sound at the current point in space.

We solve this equation using a second-order Runga-Kutta time integration scheme. The first half-step we update our state vector at time $t$ and cell $n$ as
\begin{equation}
\label{eq:TVD_halfstep}
    \mathbf{U}_{n}^{t+\Delta t/2} = \mathbf{U}_{n}^t - \left(\frac{\mathbf{F}_{n+1/2}^t-\mathbf{F}_{n-1/2}^t}{\Delta x} \right) \frac{\Delta t}{2},
\end{equation}
where we define our fluxes at each cell boundary as $\mathbf{F}_{n+1/2}^t = \mathbf{F}_{n+1/2}^{R,t} - \mathbf{F}_{n+1/2}^{L,t}$. We then recalculate the boundary fluxes, i.e. $\mathbf{F}_{n}^{t+\Delta t /2}$, and perform a full time step as
\begin{equation}
\label{eq:TVD_fullstep}
    \mathbf{U}_{n}^{t+\Delta t} = \mathbf{U}_{n} - \left(\frac{\mathbf{F}_{n+1/2}^{t+\Delta t/2}-\mathbf{F}_{n-1/2}^{t+\Delta t/2}}{\Delta x} \right) \Delta t.
\end{equation}

\subsubsection{Calculating Boundary Fluxes to Second Order}
In order to evaluate the terms in Eq. \ref{eq:TVD_halfstep} and \ref{eq:TVD_fullstep}, we need to calculate the fluxes at cell boundaries based on those at cell centers. For this we use a second order upwinding scheme which has been demonstrated to be stable and able to capture complex phenomena like astrophysical shocks.  To first order we can describe the flux at the right boundary of cell $n$ as,

\begin{equation}
   \mathbf{F}_{n+1/2}^t =  \left\{
\begin{array}{ll}
      \mathbf{F}_n^t& v > 0\\
      \mathbf{F}_{n+1}^t & v\leq 0 \\
\end{array} 
\right.
\end{equation}
While the first order scheme is conservative and doesn't produce spurious oscillations, it is highly diffusive. To this we add a  nonlinear second-order accurate total variation diminishing (TVD) scheme on top of the standard first order upwinding, for the left and right moving waves,
\begin{equation}
    \Delta \mathbf{F}_{n+1/2}^{L,t} =  \frac{\mathbf{F}^t_{n}-\mathbf{F}_{n-1}^t}{2}
\end{equation}
\begin{equation}
    \Delta \mathbf{F}_{n+1/2}^{R,t} =  \frac{\mathbf{F}^t_{n+1}-\mathbf{F}_{n}^t}{2}
\end{equation}
The correct second order correction depends on the the direction of flow and we therefore use a flux limiter, $\phi$, to determine which second order approximation is appropriate \citep{hirsch1990numerical}, i.e. 
\begin{equation}
    \Delta \mathbf{F}_{n+1/2} = \phi(\Delta \mathbf{F}_{n+1/2}^{L,t},\Delta \mathbf{F}_{n+1/2}^{R,t}).
\end{equation}
It is important to note that certain common choices for the TVD flux limiter (such as the ``minmod" limiter) would be difficult to implement differentiable due to discrete changes in sign that could lead to vanishing derivatives and poor performance in downstream inference tasks. We instead use an altered Van Leer limiter \citep{van1974towards} which uses the harmonic mean between the left and right corrections
\begin{equation}
    \text{vanleer}(a,b) = \frac{2ab}{a+b + \epsilon},
\end{equation}
where $\epsilon$ is a small correction to maintain analytical derivatives near $a=b=0$.

To go to three dimensions, we can follow the the Strang splitting scheme \citep{strang1968construction}, wherein we solve each direction sequentially and for the next timestep we reverse the order, permuting the dimensions every other timestep (i.e. XYZZYX, ZXYYXZ, YZXXZY). This splitting scheme has a number of computational/implementation advantages, as each direction is a combination of linear operators and is straightforward to differentiate. In addition, within the \texttt{JAX} environment, we can compile this the dimensional relaxation operator as a function of dimension allowing rapid evaluation on GPU.

\subsubsection{Time-stepping}

A critical choice for any simulation is how to choose a time-step. We do not implement any time-subcycling and use global time-steps. There are a two main time scales of interest, the dark matter gravitational timescale and the hydrodynamical timescale. In practice, since we are using particle mesh dynamics for the dark matter, we do not resolve viralization of the dark matter particles and velocities stay comparatively small. This results in a comparatively long timescale throughout the simulation run.

To inform our choice of timestep from the hydrodynamical physics, we use the CFL (Courant-Friedrichs-Lewy) condition. It essentially states that information cannot propagate numerically faster than the underlying physics would allow. In practical terms, a fluid element should not travel more than one grid cell during one timestep. This is expressed mathematically as the CFL coefficient C, where:
\begin{equation}
    C = \frac{v\Delta t}{\Delta x} \leq C_{max}.
\end{equation}
Here, v is the characteristic velocity (which could be the fluid velocity, sound speed, or any other relevant wave speed), $\Delta t$ is the timestep, and $\Delta x$ is the grid spacing. Typically, $C_{max}$ is set to $\leq$1 for explicit time integration schemes, and in practice, values of 0.2-0.5 are often used for good accuracy and stability margin. In our case, extra care needs to be taken in choice of this number since ensuring solution accuracy is not sufficient for ensuring derivative stability. For most applications, including parameter inference within HMC, noisy gradients can be tolerated up to a point. While there does exist some theoretical work on how to best control this noise \citep{chen2014stochastic}, in practice we can test our gradients before the inference step.

For hydrodynamical simulations, one must consider all relevant wave speeds. In the case of compressible flow, such as for cosmic baryons, this means taking into account both the bulk fluid velocity and the sound speed. The timestep must satisfy the CFL condition for the maximum wave speed: 
\begin{equation}
    \Delta t \leq \frac{(C_{max} \Delta x)}{|v| + c_s},
    \label{eq:dt}
\end{equation}
where $c_s$ is the local sound speed. This ensures that acoustic waves, shocks, and gas advection are all properly resolved in the numerical solution. 

%While Eq. \ref{eq:dt} ensures numerical stability of the forward pass, derivative calculations may require a smaller time-step in order to reach the desired stability. For most applications, including parameter inference within HMC, noisy gradients can be tolerated up to a point. In practice, one may have to tune 

Depending on application, care needs to be taken in how the number of iterations and choice of time-step is chosen in order to get to the target redshift. While one can back-propagate through our timestep calculation and enforce that the total integrated time results in the target redshift, in practice it simpler to predefine our time-stepping scheme even if it might mean excessive iterations. For constructing differentiable summary statistics, this is a fairly natural choice as one can use the first simulation to establish the time-step schedule. Significant deviations from the fiducial model may require readjusting the time-step schedule; i.e.~if structure forms at a significantly different redshift where the time-steps are too large numerical accuracy will be lost. 

Time-step accuracy can be checked explicitly by calculating $\Delta t$ with Eq.~\ref{eq:dt} for each time-step of the solution simulation and comparing it to that of the fiducial model. Since a comparatively small $C_{max}$ is used for the fiducial model only require a condition that $\Delta t_{i,\text{fid}} / \Delta t_{i,\text{sol}} \leq C_\text{fid} / C_\text{forward}$ for all time-steps $i$, where $C_\text{forward}$ is the maximum CFL number which still holds numerical accuracy on the forward pass ignoring accuracy of derivative calculations. $C_\text{forward}$ can be found via standard convergence analysis, and in our examples is usually $\sim 1.5 C_\text{fid}$.

\subsubsection{Sedov Blast Wave }
\label{subsec:sedov}

To test the forward evolution of our hydrodynamics solver in isolation we use the Sedov–von Neumann–Taylor \citep{von1941point,1946sedov,1950Taylor}  blast-wave (henceforth ``Sedov"). The Sedov test is a rigorous and demanding evaluation for any three-dimensional Eulerian or Lagrangian hydrodynamic code. In this test, a simulation box is initialized with a uniform medium of constant density and negligible pressure. A concentrated source of thermal energy is introduced at the center at time $t=0$. The main challenge lies in accurately simulating the strong spherical shock-wave that moves outward, sweeping material as it propagates through the surrounding medium. This setup is frequently used in astrophysics to model supernova explosions and the evolution of their remnants \citep{1992pavi.book.....S,blondin1998transition}. A major advantage of the Sedov blast wave problem is that it can be solved analytically \citep{landau1987fluid,sedov2018similarity} and the solution has a self similar nature. The outward propagating shock in an ideal gas has radius 
\begin{equation}
    r_{sh}(t) = \phi_0 \left( \frac{E_0t^2}{\rho_1}\right)^{1/5},
    \label{eq:sedov}
\end{equation}
where $\phi_0 = 1.15$. 

\begin{figure}
    \centering
    \includegraphics[width=0.890\linewidth,trim={0.25cm 0.3cm 0.25cm 0.25cm},clip]{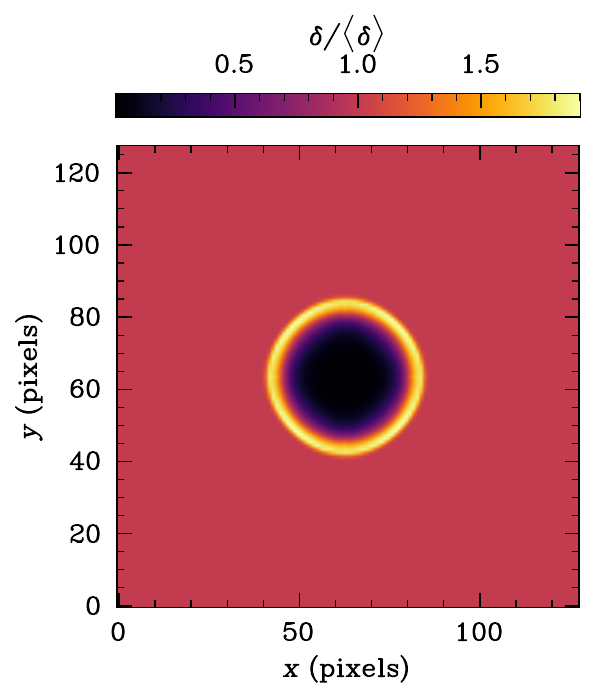}
    \caption{Central slide of a Sedov-Taylor blast-wave test conducted in a box with $128^3$ cells after 30 time steps. Initial energy to density ratio tuned for the analytical radius to be $r_{sh}(t_f)=20.0.$}
    \label{fig:sod}

    \centering
    \includegraphics[width=0.90\linewidth,trim={0.25cm 0.3cm 0.25cm 0.25cm},clip]{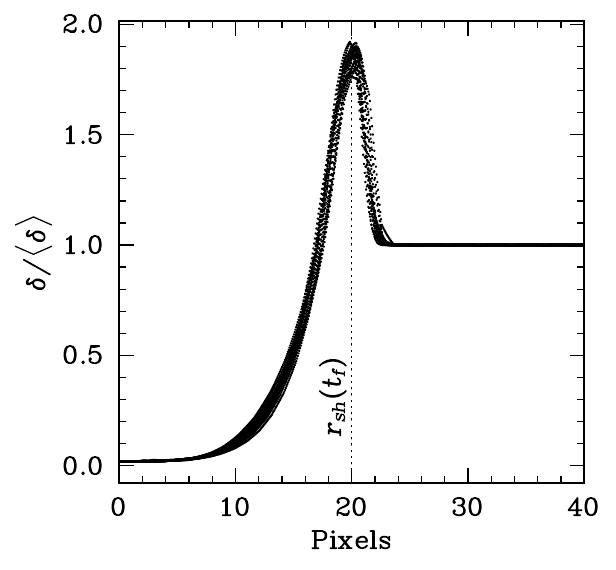}
    \caption{Density profiles at random angles from the center of the blast wave. Analytical radius from Eq. \ref{eq:sedov} shown.}
    \label{fig:prof}
\end{figure}

In our example, we evolve the initial position and energy of a Sedov Blast to make observations of the late time density field. For our configuration, we use a box with $128^3$ cells and constant initial density $\rho_1 =1$. At the initial time, $t=0$, we inject energy $E_0$ into the central pixel of our grid. The simulation is stopped after thirty timesteps. We show the central slice through our blast wave in Figure \ref{fig:sod}, as well as the variation of the radial profile in Figure \ref{fig:prof}. 
We find our results in-line with the analytical prediction, with a shock-front well resolved within roughly 2 pixels, and angular variation of the shock profile of less than 1 pixel. These results are in line with other simulations of the Sedov blast, such at those in \citet{2003PASP..115..303T}.

%We use the full three dimensional temperature field to perform a likelihood analysis for the best fit parameters, $E_0$ and $\mathbf{X}$. To the observed temperature field, we add a small uniform noise ($\Delta E = 1.0$) We use a mean squared likelihood and a uniform prior for the position in the central fifth, as well as $0<E_0<10^6$.

\begin{figure*}
    \centering
    \includegraphics[width=1\linewidth]{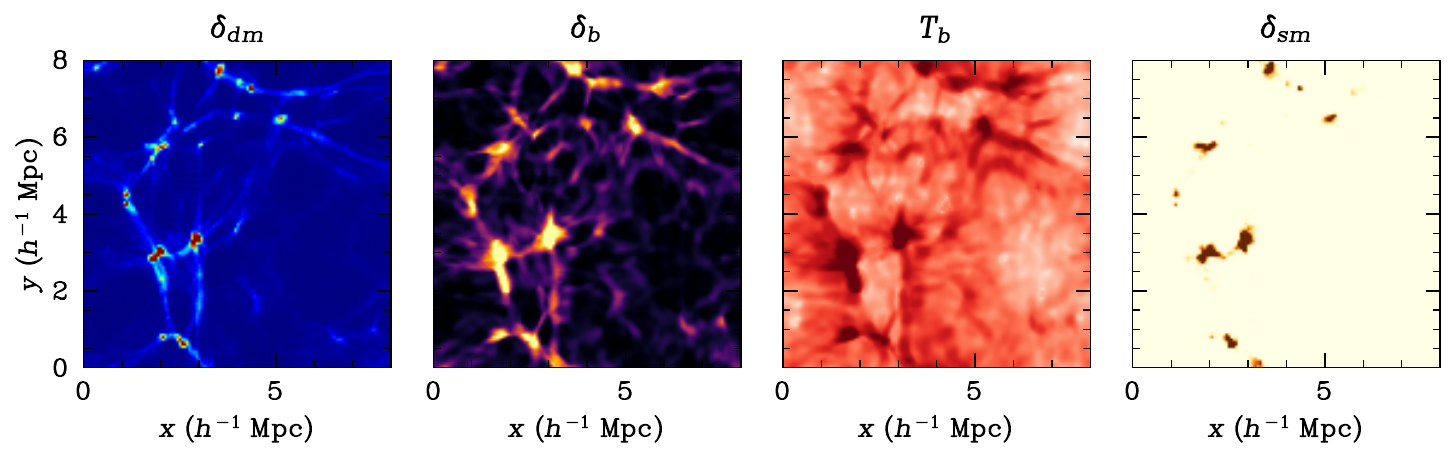}
    \caption{Example simulation in a $256^3$ resolution box evolved to $z=2.5$. Shown is the dark matter density field, baryon density field, temperature field, and the tracked star particles.}
    \label{fig:example_box}
\end{figure*}

\begin{figure}
    \centering
    \includegraphics[width=0.90\linewidth]{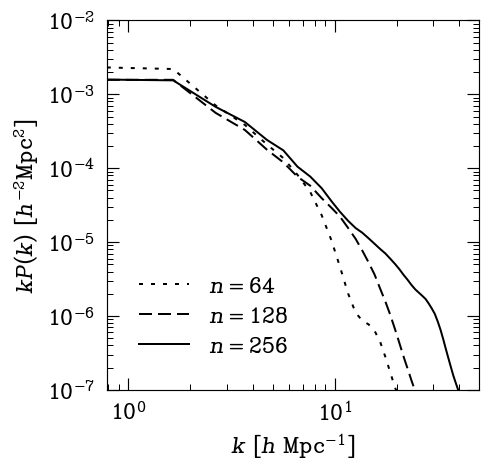}
    \caption{Convergence test of \diffhydro\ (without sub-grid physics) for evolved baryon density field for an 8 $h^{-1}$ Mpc box at $z=2.5$, changing box size while keeping the same initial conditions. We find excellent agreement even at relatively coarse resolutions up to $k\sim 7.0$ $h$ Mpc$^{-1}$.}
    \label{fig:Convergence}
\end{figure}

\subsubsection{Implementation and Differentiation}
\label{subsubsec:Implementation}

Our code, \diffhydro\ is written entirely in Python, allowing rapid development, while also utilizing \texttt{JAX} \citep{2021ascl.soft11002B} for automatic differentiation and GPU-enabled computation.  \texttt{JAX} provides familiar \texttt{NumPy}-like syntax while introducing powerful capabilities for gradient-based optimization, parallel computing, and just-in-time (JIT) compilation. This compilation is a computational technique that enables the conversion of Python functions into optimized machine code via the XLA (Accelerated Linear Algebra) compiler, enhancing computational efficiency on modern hardware. For cosmological hydrodynamical simulations, which require extensive numerical computations to model fluid dynamics and gravitational interactions across vast time steps and scales, JIT compilation offers substantial performance benefits. By reducing function call overhead and enabling optimizations such as loop fusion and vectorization, JIT compilation allows for more efficient utilization of GPUs. This efficiency can facilitate faster and more scalable simulations, enabling the high-fidelity modeling required for studying the evolution of cosmic structures.

The core innovation of \texttt{JAX} is its ability to automatically differentiate \citep{wengert1964simple, bartholomew2000automatic,griewank2008evaluating} through complex linear algebra operations and associated algorithms. This is achieved through its function \verb|jax.grad|, which computes gradients of scalar-valued functions, and \verb|jax.jacobian|, which computes full Jacobians. Unlike ``traditional" autodiff libraries (such as \texttt{TensorFlow}) that maintain a computational graph during runtime, \texttt{JAX} adopts a "trace-and-transform" approach. It traces the Python function into a functional, stateless representation and then applies transformations like differentiation (\verb|grad|), vectorization (\verb|vmap|), and parallelization (\verb|pmap|). This design makes \texttt{JAX} more composable and efficient for large-scale computations.\footnote{In practice this trade-off involves a relatively lengthy compilation step once a model is built for function values and gradients, but very rapid function evaluation. For the applications shown in Sec. \ref{sec:results}, the compilation time was approximately 10 minutes and evaluation was \~10 seconds. This rapid evaluation is key for techniques like Hamiltonian Monte Carlo.}

\texttt{JAX}'s autodiff system is based on forward-mode and reverse-mode autodiff, with a focus on reverse-mode differentiation for most deep learning/high dimensional optimization applications. Reverse-mode autodiff is particularly efficient for scalar output functions with high-dimensional inputs, such as the loss function of field level inference problems. \texttt{JAX} accomplishes this by leveraging operator overloading, a technique where every mathematical operation is wrapped in a special ``primitive" that tracks inputs, outputs, and computational rules. When \verb|jax.grad| is applied, it ``replays" these operations in reverse to compute gradients.

Unlike symbolic differentiation, which works with algebraic expressions, or finite differences, which approximates derivatives numerically, \texttt{JAX} operates directly on the computation's underlying primitives. Each primitive corresponds to a low-level operation (like \verb|add|, \verb|multiply|, or \verb|dot|), and \texttt{JAX} defines the derivatives for these primitives explicitly. As \texttt{JAX} traces the execution of a function, it records how these primitives are composed. During the backward pass, it uses the chain rule to propagate gradients from the output back to the inputs. This system enables efficient differentiation through linear algebra, such as the TVD numerical routine described above.

One concern with automatic differentiation with complex forward models is memory costs of the intermediate states and how to best balance memory usage and performance. \texttt{JAX} checkpointing is a tool that saves intermediate states (or "checkpoints") of a computation, allowing \texttt{JAX} to recompute only certain parts of the computation graph during backpropagation, rather than holding the entire graph in memory. This trade-off reduces memory usage at the cost of additional computation time. This is particularly useful in recurrent computations or large transformations, where memory constraints are significant. By strategically placing checkpoints, one can control the balance between memory usage and computation overhead. In our case, we manually specify check-pointing every 6 time-steps, i.e. after every iteration of the Strang-splitting scheme, which strikes a good balance between memory and performance. 

We show the performance of our model in Figure \ref{fig:Convergence} up to a grid resolution $256^3$. The current code is a proof of concept and not currently parallelized to be deployed across multiple GPUs, so the forward pass is memory constrained. However, even at these modest resolutions we find consistent modeling well into the non-linear regime. A full comparison to existing cosmological hydrodynamical simulations will be explored in future work with a parallelized implementation of \diffhydro.

\subsection{Subgrid Example: Star Formation and Energy Injection}
\label{subsec:subgrid}

There are many possible ways to include physics below the resolved physical scales, such as star formation and AGN activity, known as ``sub-grid physics." In this work we will focus on showing that a fairly simple stochastic subgrid-physics model which includes many features of those found in literature. In our current proof of concept, our resolution is not sufficient for full reproduction of the more complex aspects of these models which we leave for future work.

\subsubsection{Toy Model}

We focus on recreating a simplied version of the approach used in the IllustrisTNG/AREPO galaxy formation model \citep{2003MNRAS.339..289S,2010MNRAS.401..791S,2013MNRAS.436.3031V,2014MNRAS.445..175G,2017MNRAS.465.3291W,2018MNRAS.473.4077P,2020ApJS..248...32W}, which is the basis of a large number of hydrodynamical simulation suites \citep{2021ApJ...915...71V,2023MNRAS.524.2539P,2023MNRAS.521.4356X,2024arXiv240919047B}. This model is inspired by the Kennicutt–Schmidt law \citep{1959ApJ...129..243S,1989ApJ...344..685K} relating surface gas density and star formation rate.

In particular, we say there is some critical temperature, $T_c$, below which star formation is possible if a gas density critera, $\rho_c$, is exceeded. The probability of formation is proportional to a rate density, $\Pi_0$. Since, in this work, we are primarily interested in the IGM physics as opposed to the stellar populations, we are primarily concerned with the effects of the resulting supernova heating on the gas. While in principle there is a delay between star formation events and their associated supernova, this scale is quite small relative to our timestepping, self-regulation establishes itself quickly, and the delay can be neglected. We assume that the supernovae feedback energy directly heats the surrounding gas density at an equivalent energy, $E_{0}$. In practice, this creates a Sedov-Taylor blast wave, which is a common model used for supernova studies \citep{blondin1998transition,2022ApJS..262....9O}, and which we show in isolation without large scale structure and gravity in Sec. \ref{subsec:sedov}.

\begin{figure}
    \centering
    \includegraphics[width=1\linewidth]{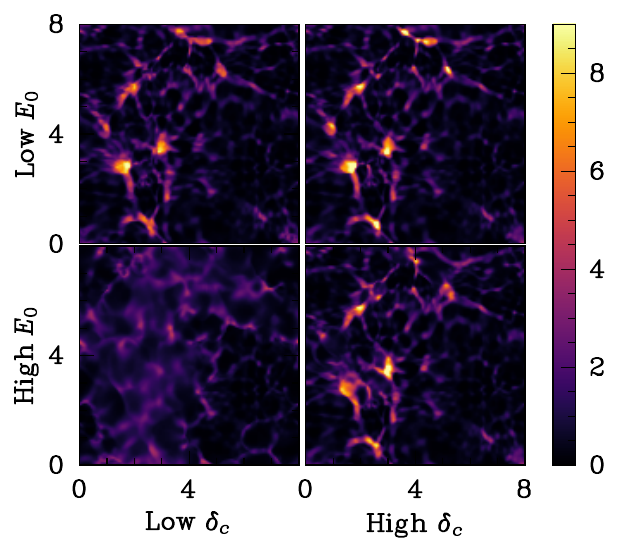}
    
\includegraphics[width=1.0\linewidth]{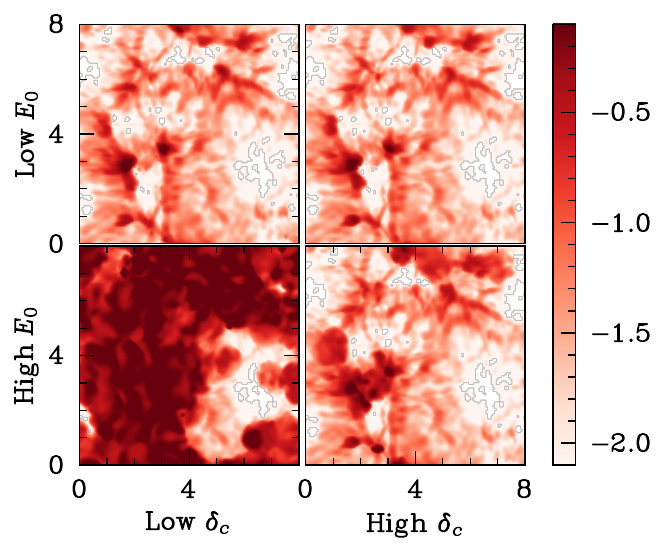}

\includegraphics[width=0.98\linewidth]{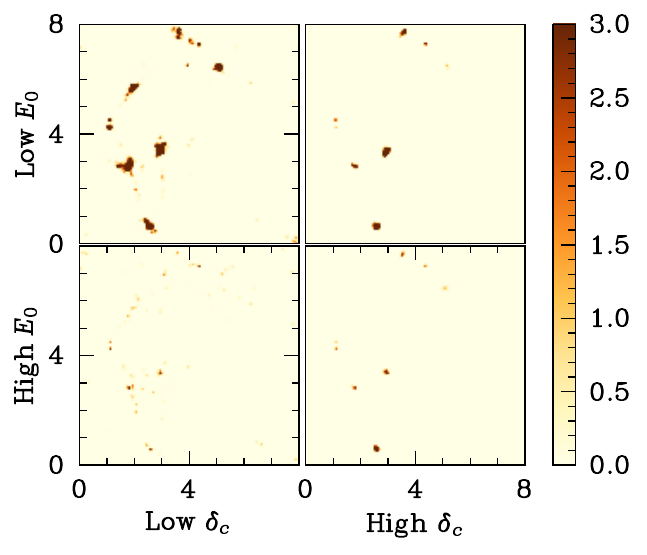}
    \caption{Demonstration on how baryon density, internal energy, and stellar field can change with varying the subgrid physics parameters $\delta_c \in [0.5,1.5]$ and $E_0 \in [0.01,0.05]$ with fixed initial conditions at $z=2.5$. Slices are projected over a 100 $h^{-1}$ kpc thick slab.}
    \label{fig:subgrid}
\end{figure}
While relatively simplistic, this model contains many of the key features of a wide variety of sub-grid physics feedback models. In particular it stochastically injects energy into the surrounding gas with some characteristic timescale. It is described by a relatively small number of parameters ($T_c$, $\rho_c$, $\Pi_0$, and $E_{0}$) which can be tuned to reproduce various observed statistics. Since we do not currently have implemented the metal heating/cooling, we need to use a high value of $T_c$ and/or a low value of the $\delta_c$ to allow star formation, however star formation will self regulate with a sufficiently large $E_0$ value.

Once star particles are created, we associate them with the nearby dark matter particles to track their future evolution. To do this we apply a cloud in cell readout routine to the 3d star field mesh at the dark matter particle positions. This provides a remapping from the Eulerian coordinate-defined stellar field to the dark matter particles. We then add this quantity to the dark matter particle's auxiliary field variable. Currently this does not affect the resulting evolution and is used for diagnostic/plotting purpose only, however additional sub-grid physics prescriptions could be added for continued IGM heating. 

We note that this approach differs from that in the standard AREPO galaxy formation models where new star particles are created and then evolved. However, once created both star particles and dark matter particles evolve as collision-less gravitationally interacting particles (i.e. via Eq. \ref{eq:dm_vlasov}) and will have the same trajectories in phase space. Since, by construction, stars form in dense environments there will likely be a dark matter particle close to the formation zone and will be a suitable choice as a sample for the phase space evolution. Additional care may be needed to properly account for differences in gas and dark matter velocity depending on application.

We show one example including the star field in Figure \ref{fig:example_box}. This box is evolved until $z=2.50$ and shows the dark matter field, the gas density, gas temperature, and stellar density field. As expected, we find the baryons trace the dark matter structures and the stellar field is a highly biased version of the baryon field.

We show examples of our varying our subgrid physics model in Figure \ref{fig:subgrid}, evolved to $z=2.0$. In extreme cases, our feedback is able to completely overwhelm the gravitational binding energy and disrupt the baryon structure, an extreme analogy of a supernova event. At more reasonable levels of feedback, we find the expected result of heating the IGM in dense regions. As the critical density, $\delta_c$ increases, we find less impact of our feedback since the large scale structure takes longer to evolve to this density and there will be fewer star formation events.

%We can therefore write the energy change per cell with star formation as
%\begin{equation}
%    \frac{d}{dt} (\rho E) = \beta u_{SN} \frac{\rho_c}{t_*} 
%\end{equation}
%where $\beta$ is the ``instantaneous" supernova efficiency rate (i.e. assuming the stars supernova at a timescale below our simulation timestep), $u_{sn}$ is the supernova energy density

\subsubsection{Differentiation}

Our goal is to compute gradients of any observable with respect to sub-grid physics parameters throughout the simulation. Conventional understanding suggests that differentiating through stochastically sampled discrete random variables, like the formation of an AGN or star cluster, is not feasible. However, advances in Reinforcement Learning have introduced techniques to handle such discrete variables in the training of deep neural networks via backpropagation. 

Specifically, we apply the Gumbel-Softmax or CONCRETE method, which uses continuous distributions to approximate the sampling process of discrete stochastic variables, such as galaxies, in a differentiable manner \citep{2016arXiv161101144J,2016arXiv161100712M} . This approach leverages two key insights: 1) a re-parameterization of a discrete (or categorical) distribution using the Gumbel distribution, and 2) making the corresponding function continuous by applying a continuous approximation controlled by a temperature parameter, which, at zero temperature, reduces to the original discontinuous form. This approach has been used to populate discrete galaxies into halos (i.e. a halo occupancy distribution) to allow differentiation of galaxy power-spectra with respect to the HOD parameters \citep{2024diffhod}.

The central concept of this technique, known more generally as the reparametrization trick, is to express samples $z$ from a given parametric distribution $\mathbb{P}_{\theta}$ as a deterministic, differentiable transformation $f$ applied to a fixed distribution $\mathbb{P}_{\epsilon}$: 
\begin{equation} z = f(\theta, \epsilon) \quad \text{where} \quad \epsilon \sim \mathbb{P}_{\epsilon} \end{equation} 
This reparameterization of the samples bypasses the need to differentiate the stochastic variable $\epsilon$ when computing the derivatives of some downstream function $h$ with respect to the distribution parameters $\theta$. This can be written in terms of the expectation with respect to the random variable $\mathbb{E}_{z}$ as: 
\begin{equation}
\label{eq:reparam}
    \frac{\partial}{\partial \theta} \mathbb{E}_{z \sim \mathbb{P}_\theta}\left[ h(z) \right] = 
    \mathbb{E}_{\epsilon \sim \mathbb{P}_\epsilon} \left[ \frac{\partial}{\partial \theta} h(f(\theta, \epsilon)) \right] %= \mathbb{E}_{\epsilon \sim p_\epsilon} \left[ \frac{\partial h}{\partial f}  \frac{\partial f}{\partial \theta} \right].
\end{equation}
On the right-hand side of this equation, the derivative now only involves taking the gradient of a deterministic function of $\theta$, with $\epsilon$ treated as an input to the function. While this can be used to construct real valued outputs, we need our mapping to be discrete (either star particle is formed, or not). We can then use the Gumbel trick \citep{gumbel1954statistical} combined with a softmax formula to construct an array $z$ to describe the star forming field per timestep;
\begin{equation}
\label{eq:gumbelsoftmax}
    \hat{z} = \frac{\exp((\log(\pi_0)+g_0)/\tau)}{\exp((\log(\pi_0)+g_0)/\tau)+\exp((\log(1-\pi_0)+g_1)/\tau)},
\end{equation}
where $g_i$ are independent random variables drawn from the Gumbel distribution between 0 and 1, $\textrm{Gumbel}(0, 1)$, which are treated as inputs to our model (see Eq. \ref{eq:reparam}). 

In the limit of $\tau \rightarrow 0$, Eq. \ref{eq:gumbelsoftmax} will exactly match the true discrete random field, however we can differentiate this formula in relation to the class probabilities $\pi$ if $\tau > 0$. Care needs to be taken to maintain numerical accuracy, as very small values of $\tau$ will result in significant noise due to the divisions of large numbers in Equation \ref{eq:gumbelsoftmax}. In practice, trial and error is the most expedient way to choose $\tau$ that results in the right stochastic properties while maintaining sufficient numerical accuracy (see \citet{2024diffhod} for more discussion). For examples in this work, we found $\tau = 0.1$ suitable.

In order to maintain smooth derivatives and improve general performance, we use a sigmoidal activation, $\sigma(x) = (1+ \exp (-x))^{-1}$, for the temperature and density thresholds instead of a sharp cutoffs. We define our class probability to be 
\begin{equation}
    \pi_0(\rho,T) = \Pi_0 \sigma\left(\frac{\delta - \delta_c}{a}\right) \sigma\left(\frac{T_c - T}{b}\right),
\end{equation}
where $a$ and $b$ describe how sharp the transition is near the critical temperature. In the limit as $a,b \rightarrow 0$ we recover the discontinuous transitions. In principle, these could be treated as additional sub-grid physics parameters and jointly fitted, but in this work we treat them as hyper-parameters (like $\tau$) which are set at values to ensure numerical stability during backpropagation.

\section{Applications}
\label{sec:results}
In this section we highlight two applications of our differentiable framework; 
\begin{enumerate}
    \item Jointly sampling cosmology and subgrid physics from mock cosmological summary statistics
    \item Field level inference of initial matter density field given noisy mock observations
\end{enumerate}

%\subsection{Posterior Inference on Polytropic Index from Merger Event}

%For a more complex example, we model a merger event between two large gas clouds, inspired by the globular cluster star merger modelled in \citet{2003PASP..115..303T}. For this work we consider an off-axis collision of two gas masses, with $M = 0.8 M_\odot$ and $R= 0.955 R_\odot$. We model the interior pressure as $P \propto \rho^{1+1/n},$ with polytropic index $n=3$. We initialize our gas spheres on zero-energy parabolic orbits with a pericenter separation of $0.24 R_\odot$.

%We perform our simulation in a box with $256^3$ cells and show four time slices in [FIGURE].

%For our inference problem, we construct a mock observation of the 

\subsection{Joint Stochastic Subgrid and Cosmological Parameter Inference from Baryon Power Spectra}

\begin{figure}
    \centering
    \includegraphics[width=1.0\linewidth]{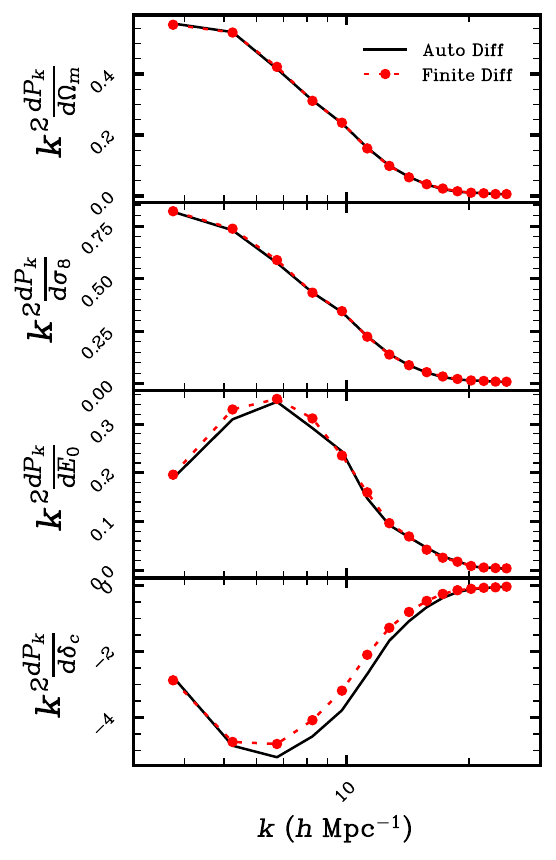}
    \caption{Comparison of our automatically calculated power spectra derivatives vs. those found via finite differences at $z=2.97$. We perform our comparison at a fixed realization in order to show the accuracy of our derivative calculation. While methods are susceptible to numerical noise, downstream likelihood calculations aren't significantly effected. Note that at these cosmological values, $dP_k/d\Omega_m$ and $dP_k/d\sigma_8$ have very similar shapes, but their ratio does have a k-dependence.}
    \label{fig:pk_deriv}
\end{figure}

\begin{figure*}
    \centering
    \includegraphics[width=0.8\linewidth]{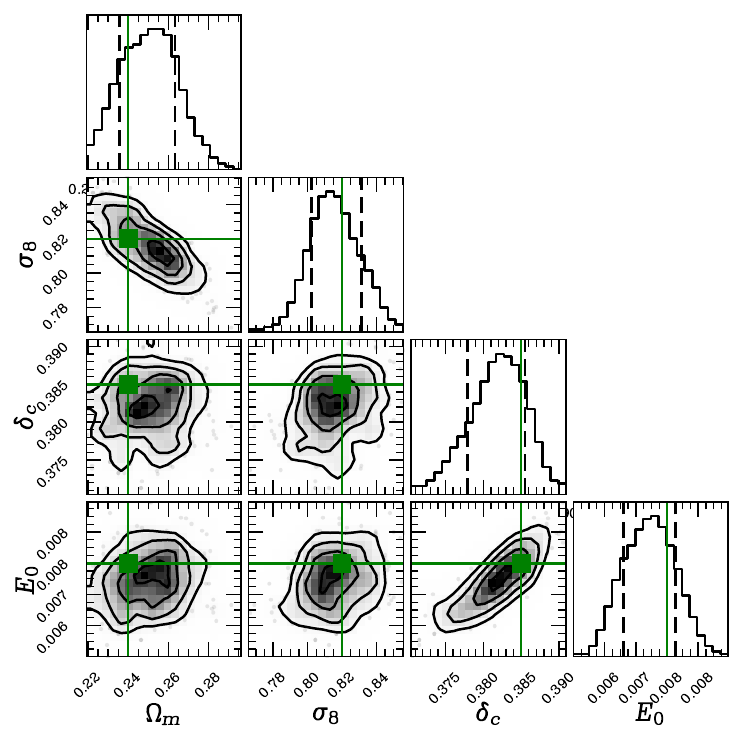}
    \caption{Hamiltonian Monte Carlo chain estimating the posterior for cosmological and astrophysical parameters given late-time baryon power spectra. The green point is the true simulated value, which is within one sigma of the estimated distribution mean for all four varied parameters.}
    \label{fig:corner}
\end{figure*}
For our first example, we will construct a mock observation of the baryon density power spectra including sub-grid physics, and then infer posteriors on the associated parameters. A similar approach was taken in \citet{2023A&A...679A..61L} in the context of weak lensing analysis, where differentiability was used to construct rapid Fisher forecasts for various summary statistics. In our case, we will construct the entire posterior via Hamiltonian Monte Carlo sampling of the physical parameters. 

We use our model including both baryons and dark matter to simulate a three dimensional volume. We use a particle grid resolution of $128^3$, a physical scale of $8$ $h^{-1}$ Mpc, and evolve our field to $z=3.0$. 

\begin{figure}
    \centering
\includegraphics[width=1.0\linewidth]{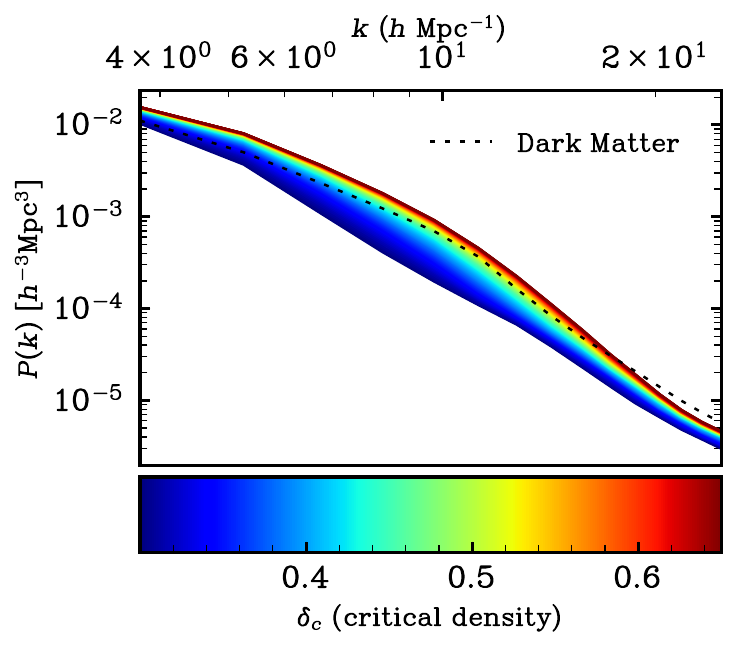}

\includegraphics[width=1.0\linewidth]{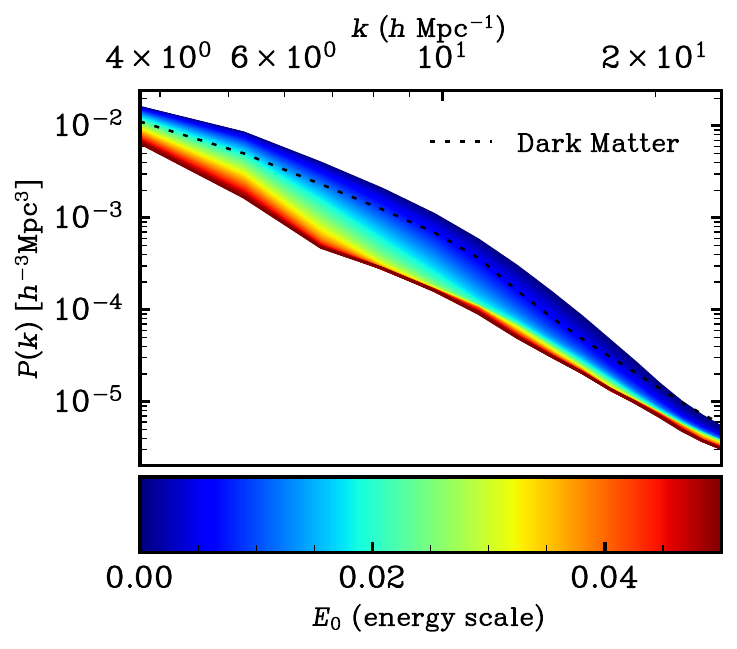}

    \caption{Effects of varying two sub-grid physics parameters at $z=2.97$ shown in relation to the dark matter power spectra at the fiducial parameters.}
    \label{fig:pk_vary}
\end{figure}

We show the effect of varying these parameters  on the power spectra in Figure \ref{fig:pk_vary}.  We study the joint inference of $E_0$, $\delta_c$, $\sigma_8$, and $\Omega_m$, holding all other parameters constant. These parameters are chosen to be representative of both cosmology and sub-grid physics, while keeping the dimensionality small for demonstration.

To model a mock observation, we evolve our system using the methods described in Sec. \ref{subsec:subgrid}, and calculate the power spectra of the baryon density using a differentiable routine described in \citep{2024diffhod}. For our likelihood covariance, we run 100 boxes with independent initial seeds and use the full covariance of those boxes baryon power-spectra.

To test our differentiable framework, we show the automatic derivatives of our power spectra with respect to the parameters in Figure \ref{fig:pk_deriv}, compared with those calculated with finite differences. We find satisfactory accuracy across a range of scales. Also note that due to the accept-reject step of HMC, inaccurate derivatives will not necessarily lead to inaccurate posteriors but might reduce acceptance rate. 

To evaluate the posteriors using our differentiable hydrocode, we sample over the joint subgrid physics and cosmological parameters using a 
Hamiltonian Monte Carlo (HMC) \citep{duane1987hybrid,neal2011mcmc}. We use the NoUTurn HMC implementation \citep{hoffman2014no} in \texttt{BlackJAX} \citep{cabezas2024blackjax} to adaptively set our step size. We use Gaussian priors on our parameters, centered at the true values, with $\sigma=0.002$ for $E_0$ and  $\sigma = 0.02$ for $\Omega_m$/$\sigma_8$/$\delta_c$.

We use one chain initialized around our fiducial cosmological and hydrodynamical parameters with 500 steps (50 steps of burn-in). We found our sampler was able to generate negative autocorrelation, with chains having an effective sample size\footnote{The effective sample size can be calculated from a Markov chain as
\begin{equation}
    N_\textrm{eff} = \frac{N}{1+\sum\limits_{t=1}^{\infty} \rho_t},
\end{equation}
where $N$ is the number of samples and $\rho_t$ is the autocorrelation
of length $t$.} of $\sim 6600$. We show the results of this chain in Figure \ref{fig:corner}. We find the well-known degeneracy in the $\sigma_8$-$\Omega_m$ plane, as well as a degeneracy in the $\delta_c$-$E_0$. This latter degeneracy is due to the need for additional energy to compensate for a higher critical density, and therefore fewer possible star formations. Due to the small box size, we are unable to recover possible degeneracies between the sub-grid physics parameters and the cosmological parameters.

\subsection{Field level inference}
\label{subsec:FLI}

For our second example, we aim to reconstruct the full initial conditions corresponding to the observed volume. While in full generality these conditions should be sampled jointly with the model parameters, in this example will aim to reconstruct the conditions at fixed cosmology. In addition, for this example, we turn off the subgrid physics model to avoid having to jointly sample the stochastic field. 

We construct a mock observable, $\mathbf{d}$ that is proportional to gas-mass-weighted 3-dimensional temperature, i.e. a noisy measurement of $\delta_b T$. This corresponds roughly to the internal energy density of the field and is meant to represent a generic IGM/ICM probe. We apply Gaussian pixel noise with two different models. For our high S/N case, we apply noise with variance equal to the data variance, i.e. S/N = 1. For our low S/N case, we apply noise with 5 times the data variance, i.e. S/N = 5. 

Following the approach in \citep{2019TARDISI,2021TARDISII}, we perform an optimization to infer the initial conditions, $\mathbf{s}$, rather than sampling for computational expediency. In addition, in many cases variations around an optimized solution can be used for constructing robust posteriors with far fewer function evaluations than full sampling \citep{2019Horowitz,2022PhRvD.105j3531M}. Our log-posterior is the sum of a noise-weighted mean squared error over the pixels and a prior term, i.e.
\begin{equation}
    \log\mathcal{L} = \frac{1}{2} \sum_i \left[\frac{(d_i-F(\mathbf{s})_i)^2}{N_i} + |\tilde{s}_i|^2\right],
\end{equation}
where $F$ is \diffhydro, $N_i$ is the variance of the pixel noise in the $i$th pixel. The second term is our prior, in this case we aim to minimize the initial phases, $\mathbf{\tilde{s}}$ (i.e. the transfer function de-convolved initial conditions). Note that this second term is cosmological dependent, and we perform this optimization at our fiducial cosmology.

To optimize, we use the optax \citep{deepmind2020jax} implementation of the Limited-memory Broyden–Fletcher–Goldfarb–Shanno (L-BFGS) quasi-Newtonian method, with convergence criteria $\epsilon = 10^{-5}$. L-BFGS is an iterative algorithm designed for solving unconstrained optimization problems, particularly effective in high-dimensional spaces. It is a variant of the BFGS method that approximates the inverse Hessian matrix using a limited amount of memory, making it computationally efficient for large-scale problems. L-BFGS achieves this by storing only a few vectors past vectors (in our case $M=8$) that represent the gradient updates, avoiding the explicit computation and storage of the Hessian matrix. This property, combined with its fast convergence and ability to handle non-linear and differentiable functions, makes L-BFGS particularly suited for high-dimensional optimization tasks. 

\begin{figure*}
    \centering
    \includegraphics[width=0.90\linewidth]{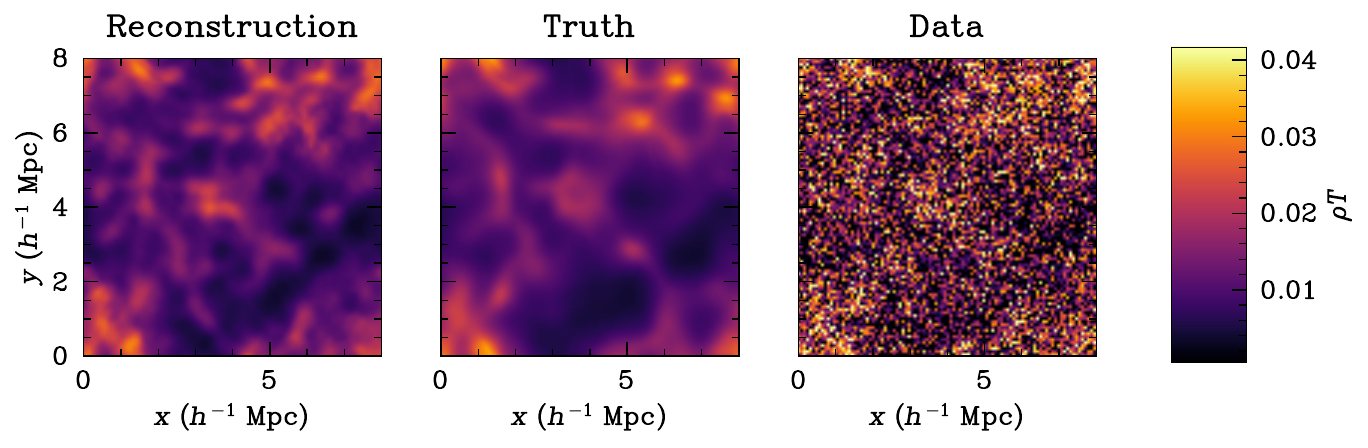}
    \includegraphics[width=0.90\linewidth]{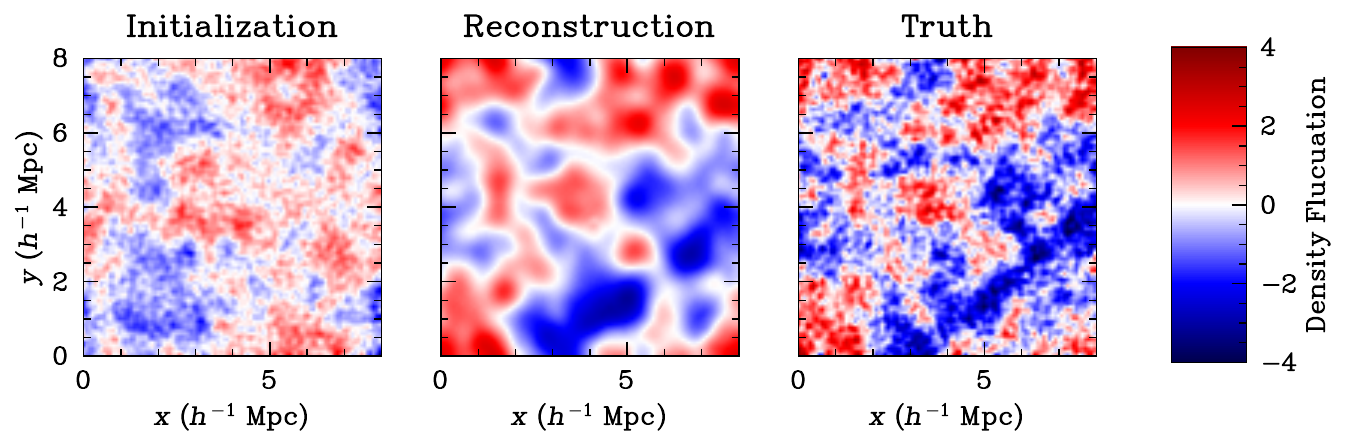}
    \caption{\emph{Top:} Reconstruction of the late time internal energy field compared to the mock true field, with the data realization in the slice for comparison. \emph{Bottom:} Reconstructed initial conditions compared to true initial condition. The initialization of the field is also shown to demonstrate our optimization started from a true random field.}
    \label{fig:fli}
\end{figure*}

\begin{figure}
    \centering
\includegraphics[width=1.0\linewidth]{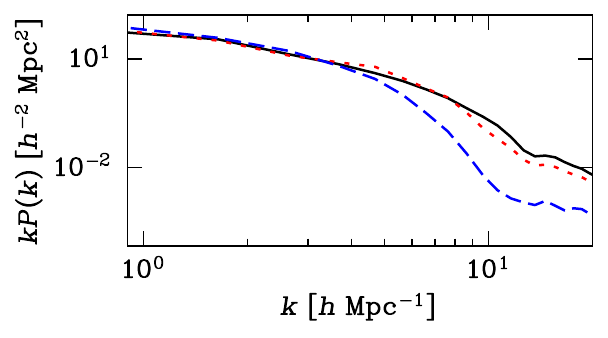}

\includegraphics[width=1.0\linewidth]{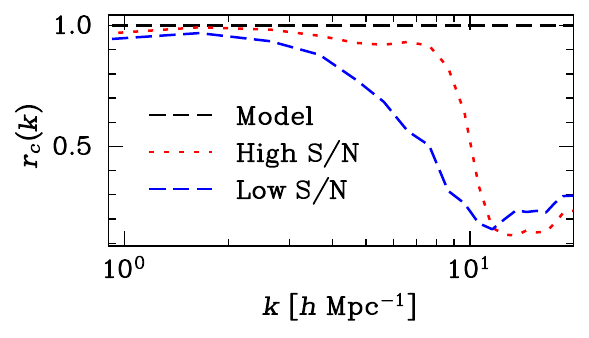}

    \caption{Powerspectra and cross correlation coefficient for our two mock samples compared to the simulated power spectra.}
    \label{fig:cross_cor}
\end{figure}

We show our qualitative results for the low S/N case in Figure \ref{fig:fli} for both the reconstructed initial conditions and the expected target redshift energy density. Small scale modes are washed out due to the pixel noise in an analogous way as a Wiener Filter (e.g. \citet{2019Horowitz}, however we are still able to reconstruct the large wavelength modes and recover the significant cosmic structure. Quantitatively, we can examine the reconstructions performance in terms of reconstructed power spectra and cross correlation coefficient. We show these quantities in Figure \ref{fig:cross_cor}. The cross correlation coefficient is a measure of how well the actual phases of model are reconstructed as a function of scale, and can be written in terms of the auto and cross power spectra, $P(k)$, as 
\begin{equation}
    r_c(k) = \frac{P_{XY}(k)}{\sqrt{P_{XX}(k) P_{YY}(k)}},
\end{equation}
where X and Y are the reconstruction and model fields respectively. As expected, we find the higher signal mock to provides a more accurate reconstruction across a range of scales, as well as a more accurate reconstructed power spectra.

\section{Conclusions}

In this work we have demonstrated a concept of a fully differentiable hydrodynamical simulation, applicable for a range of physical problems. Our approach solves the underlying Euler equations without emulation or other approximate techniques and allows for inclusion of a range of sub-grid physics prescriptions. We have demonstrated its applicability to both summary statistic based parameter inference and field-level reconstructions.

Below we list some possible extensions of our implementation of \diffhydro as well as possible alternative approaches depending on application:
\begin{enumerate}
    \item \emph{Fixed hydrodynamical grid:} 
The scales resolved in the current implementation are best suited towards modeling intergalactic medium physics, with an eye toward understanding Ly-$\alpha$ absorption. This has allowed us to work at fixed grid scales since the IGM spans vast low-density volumes and is fully absorbed near to halos where adaptive methods methods would become relevant. Broadening the application of this work to other observables may require the implementation of adaptive mesh refinement \citep{truelove1998self, norman1999cosmological} and/or nested grid methods \citep{matsumoto2003fast}. Backpropagation through adaptive refinement methods would likely be challenging algorithmically due to the complex tree-structure that emerges. It may be sufficient to generate a refinement scheme on the forward pass, and use the same scheme during back-propagation, however this will likely increase numerical sensitivity as well as require recompilation of the model at each iteration.
    \item \emph{Dark matter particle dynamics:} Related to the fixed hydrodynamical grid is the use of fixed particle mesh dynamics instead of full n-body dynamics for the dark matter evolution. This may present serious issues in resolving galaxy-scale structures, and the full impact of this effect is dependent on the end-science case. Beyond the possible approaches discussed in Sec. \ref{subsec:dm}, an exact strategy is the use of full n-body dynamics on the forward pass (i.e. running a standard hydrodynamical simulation) which will be used for the likelihood calculation, and then use \diffhydro\ for derivative information for the update in the sampler.
    \item \emph{Memory limitations with back propagation:} Memory, for both the dark matter and baryonic fields, might become a significant limitation as the simulation resolution increases. In principle, one can also use adjoint based methods (i.e. PMWD, \citet{2024ApJS..270...36L} for a particle mesh dark matter implementation), where back propagation is not used directly, but instead the gradients are propagated through the adjoint solver. While the Euler equations are well known to admit an adjoint solution \citep{jameson1988aerodynamic}, in a our context it also requires an adjoint equation for the sub-grid physics (see Sec. \ref{subsec:subgrid}), which may be impossible depending on their construction. 
\item \emph{Parallelization across GPUs}: Currently, our implementation is on a single GPU, which limits the effective resolution and/or time-range in which we can model. The core TVD scheme for gas physics ( Sec. \ref{subsec:gas}) is straightforward to parallelize across GPU by splitting up the volume across the GPUs. Communication between the GPUs will be limited to the boundaries by construction. However, gravitational dynamics are currently implemented via a mesh method requiring fast Fourier transformations over the entire volume. While distributing FFTs across multiple GPUs is not itself a fundamental limitation \citep{2015arXiv150607933G,10.1145/3539781.3539797}, extra care is needed to ensure proper backpropagation in a compiled \texttt{JAX} environment.\footnote{For one possible approach to GPU distributed FFT in \texttt{JAX}, see \url{https://github.com/DifferentiableUniverseInitiative/jaxDecomp}.}

    \item \emph{Mapping to observables:} For this work, we study quantities that are direct functions of the baryon density and temperature field. In general, additional steps are required to map to actual observables. For example, Lyman-$\alpha$ Forest is modeled by including a thermo-ionization model, an absorption profile, and an integration along the line of sight including redshift space distortions. While these sorts of complex physical processes can be included in a differentiable way \citep{2024thalas}, for the purpose of clarity we have just used the base simulation outputs and their corresponding power spectra. Other observables common in astrophysics, such as cluster counts and other observables which rely on discrete assignment, may present theoretical challenges in implementing within a end-to-end differentiable framework. 
\end{enumerate}

The novel approach of \diffhydro\ allows a range of additional applications for cosmological and astrophysical inference beyond that available to traditional techniques.  Possible applications include:
\begin{enumerate}
    \item \emph{Improving fidelity with machine learning methods:} Differentiable cosmological hydrodynamical simulation allows the implementation of various machine learning-based refinement strategies for physical simulations. For example, "solver-in-the-loop" methods integrate deep learning augmentation at each time-step to refine the solution \citep{2020arXiv200700016U}. One can also apply deep learning models afterwards for refinement of relatively coarse (un-converged) simulations \citep{2023ApJ...958...21J,2024arXiv241116920J} or constrained generation of additional fields \citep{2022hyphy} at each time-step.
    \item \emph{High dimensional sub-grid physics inference:} Differentiable hydrodynamical simulation opens the door to implementing more expressive sub-grid physics via neural network models. Network parameters can be trained based on actual data by backpropagation through both the neural network and the simulation simultaneously. This would enable very expressive sub-grid models which could include history information (i.e. previous time-slice states) and non-local information. These sorts of hybrid end-to-end ``online" training techniques have been used recently for Earth weather forecasting \citep{kochkov2024neural}. These methods could further be employed as a model-independent search for new physics.
    \item \emph{Numerical scheme for gas evolution:} 
As additional physics are added and/or resolution is increased, the numerical sensitivity of solving our equations increases due to range of time scales involved. In this regime, it may be important to go beyond our explicit hydrodynamics solver and use implicit techniques. It is straightforward to implement implicit equation solving within \texttt{JAX} since the Jacobian information can be computed via automatic differentiation (see \citet{kidger2021on} as well as earlier Julia implementations \citep{rackauckas2017differentialequations}). This approach was used to solve the Einstein-Boltzmann equations in \citet{2024JCAP...06..063H} and \citet{2023zndo..10065126L}, and could be extended to the physics studied here.
    \item \emph{Expansion to ISM/ICM physics:} The techniques developed in this work for cosmological hydrodynamical simulations could also be applied to smaller-scale simulations of the interstellar medium or intra-cluster medium. Physics in existing codes, such as Athena \citep{2008ApJS..178..137S,Stone2020} (magnetohydrodynamics), HARM \citep{2003ApJ...589..444G} (relativistic corrections), or FLASH \citep{2000ApJS..131..273F} (stellar reactions)  could be implemented within this framework. Insights from these small-scale simulations could also be used to infer physics at cosmological scales \citep{2022ApJS..262....9O,2024ApJ...975..183O}, possibly within a single inference pipeline. Modeling these effects are a key component of upcoming and ongoing IGM/ICM surveys at both low \citep{2022ApJ...928....9L,2024ApJ...973..151K} and high redshifts \citep{2022PFSGE}.
  
\end{enumerate}

%We also note that solving the Euler equations in the regimes of interest for IGM physics are not ``stiff" differential equations \citep{hairer1991ii} since the TVD scheme described is well controlled and there exists only two characteristic timescales (set by gravity and the gas sound speed). This is important as stiff equations require significant care in their time integration, including the use of a high-order implicit time integration scheme. 

\section*{Acknowledgements}
We thank Khee-Gan Lee, Kentaro Nagamine, Peter Behroozi, Jamie Sullivan, Philip Mocz and Yuri Oku for their helpful insights and comments.

This research used resources of the National Energy Research Scientific Computing Center, a DOE Office of Science User Facility supported by the Office of Science of the U.S. Department of Energy under Contract No. DEC02-05CH11231.

\bibliographystyle{mnras}
\bibliography{example,mypapers}
\end{document}